\documentclass[aps,prb,twocolumn,notitlepage,showpacs,superscriptaddress,am]{revtex4-1}%
\usepackage{graphicx}
\usepackage{amsmath}
\usepackage{amssymb}
\usepackage{color}
\usepackage{amsfonts}%
\providecommand{\U}[1]{\protect\rule{.1in}{.1in}}

\def\cm{cm$^{-1}$}

\def\stf{$\kappa$-(BEDT\--STF)$_2$\-Cu$_2$(CN)$_3$}
\def\STF{$\kappa$-[(BEDT\--TTF)$_{1-x}$(BEDT-STF)$_{x}$]$\rm _2\-Cu_2(CN)_3$}
\def\Cu{$\kappa$-(BEDT\--TTF)$_2$\-Cu$_2$(CN)$_3$}
\def\Ag{$\kappa$-(BEDT\--TTF)$_2$\-Ag$_2$(CN)$_3$}

\def\Cl{$\kappa$-(BEDT\--TTF)$_2$\-Cu[N(CN)$_2$]Cl}

\begin{document}
\title{Bandwidth-tuning from insulating Mott quantum spin liquid to 
Fermi liquid\\
via chemical substitution in $\kappa$-[(BEDT-TTF)$_{1-x}$(BEDT-STF)$_x$]$_2$Cu$_2$(CN)$_3$}
\author{Y. Saito}
\affiliation{1.~Physikalisches Institut, Universit\"{a}t Stuttgart, 70569 Stuttgart, Germany}
\affiliation{Department of Physics, Graduate School of Science, Hokkaido University, Sapporo 060-0810, Japan}
\author{R. R\"osslhuber}
\affiliation{1.~Physikalisches Institut, Universit\"{a}t Stuttgart, 70569 Stuttgart, Germany}
\author{A. L\"ohle}
\affiliation{1.~Physikalisches Institut, Universit\"{a}t Stuttgart, 70569 Stuttgart, Germany}
\author{M. Sanz Alonso}
\affiliation{1.~Physikalisches Institut, Universit\"{a}t Stuttgart, 70569 Stuttgart, Germany}
\author{M. Wenzel}
\affiliation{1.~Physikalisches Institut, Universit\"{a}t Stuttgart, 70569 Stuttgart, Germany}
\author{A. Kawamoto}
\affiliation{Department of Physics, Graduate School of Science, Hokkaido University, Sapporo 060-0810, Japan}
\author{A. Pustogow}
\affiliation{1.~Physikalisches Institut, Universit\"{a}t Stuttgart, 70569 Stuttgart, Germany}
\author{M. Dressel}
\affiliation{1.~Physikalisches Institut, Universit\"{a}t Stuttgart, 70569 Stuttgart, Germany}

\date{\today}

\begin{abstract}
The electronic properties of molecular conductors can be readily varied via physical or chemical pressure as it increases the bandwidth $W$; this enables crossing the Mott insulator-to-metal phase transition by reducing electronic correlations $U/W$. Here we introduce an alternative path by increasing the molecular orbitals when partially replacing sulfur by selenium in the constituting bis-(ethylene\-dithio)-tetra\-thia\-fulvalene (BEDT-TTF) molecules of the title compound. We characterize the tuning of the insulating quantum spin liquid state via a Mott transition to the metallic Fermi-liquid state by transport, dielectric, and optical measurements. At this first-order phase transition, metallic regions coexist in the insulating matrix leading to pronounced percolative effects most obvious in a strong enhancement of the dielectric constant at low temperatures.
\end{abstract}

\maketitle
\section{Introduction}
The fundamental nature of the Mott insulator-metal transition remains subject of controversy and debate. Early theories \cite{Mott90,Georges96,Vollhardt12} favored a robust first-order scenario but apart from hysteresis effects it has proven difficult to identify conclusive experimental evidence for the expected phase-coexistence region \cite{Imada98,Lefebvre00,Limelette03a,Limelette03b,Kagawa04,Kurosaki05}.
Alternative viewpoints envision a more continuous crossover at finite temperature featuring aspects of quantum criticality~\cite{Senthil08,Vucicevic13,Furukawa15,Furukawa18}.
Recently, quantum spin liquids were recognized as the materials
best suited for studying the pristine Mott state in absence of magnetic order~\cite{Pustogow18b}, and its low-temperature transition to a Fermi liquid  \cite{Furukawa18}.

Since most molecular solids are rather soft, their electronic properties can readily be tuned by relatively low hydrostatic pressure \cite{IshiguroBook,*ToyotaBook,*LebedBook,*MoriBook,Yasuzuka09}. Nevertheless, many experiments become cumbersome, inaccurate or even impossible when confined to pressure cells; hence alternative methods are desirable. In the case of charge-transfer salts, reducing the size of the anions acts like chemical pressure that brings the organic donor molecules closer together. This approach is frequently applied to quasi-one-dimensional TMTTF or quasi-two-dimensional BEDT-TTF salts
\cite{YamadaBook,IshiguroBook,*ToyotaBook,*LebedBook,*MoriBook}. Already in the 1970s it was realized by Bechgaard and others that replacing sulfur by selenium in the fulvalene-type central unit TTF increases the orbital overlap of adjacent molecules and thus the bandwidth; eventually leading to superconductivity in (TMTSF)$_2$PF$_6$ \cite{Jerome80}.

\begin{figure}[b]
	\centering
	\includegraphics[width=1.0\columnwidth]{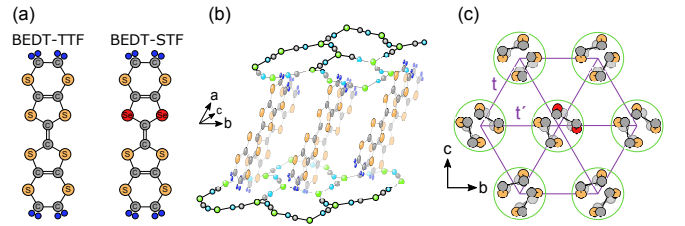}
	\caption{(a) Organic donor molecules  bis-(ethylene\-dithio)-tetra\-thia\-fulvalene, called BEDT-TTF and  bis-(ethylene\-dithio)-di\-seleniumdi\-thia\-fulvalene, abbreviated BEDT-STF. In the latter case two sulfur atoms of the inner rings are replaced by selenium. (b) The crystal structure contains dimers of the donor molecules forming layers in the $bc$-plane which are separated by the Cu$_2$(CN)$_3$ anion sheets. (c) The dimers are arranged in a triangular pattern with transfer integrals $t^{\prime}/t = 0.83$ close to complete frustration. The STF-substitution leads to a spatially random extension of the transfer integrals due to the larger molecular orbitals. }
	\label{fig:structure}
\end{figure}

Here we apply this approach to quasi-two-dimensional BEDT-TTF salts, where we replace two out of the four inner sulfur atoms by selenium as depicted in Fig.~\ref{fig:structure}(a), resulting in BEDT-STF \footnote{In contrast to the so-called BETS molecule (BEDT-TSF, bis-ethylene-dithio-tetraselenafulvalene) which contains four slenium in the fulvalene unit.
H. Kobayashi, H. Cui, and A. Kobayashi, Chem. Rev. {\bf 104}, 5265 (2004). }. While \Cu\ is a Mott insulator, known as a quantum spin liquid, \stf\ is a good conductor in the entire temperature regime.
Performing the substitution on the donor molecules ensures that the polymeric anion network and the crystal symmetry remain unaffected, which is crucial for the highly frustrated compounds.
This way we can cross the Mott transition from a paramagnetic insulator -- that is a quantum spin liquid --
to a paramagnetic metal, without any indications of magnetic order down to lowest temperatures.

In the present study we focus on the electronic properties and investigate the coexistence regime at the first-order phase transition by transport, optical and dielectric methods. We observe a divergency in the permittivity as the hallmark of a percolative transition; in other words,
we prove the insulator-metal phase coexistence below the critical endpoint $T_{\rm crit} \approx 15$-20~K.

\section{Experimental Details}
Single crystals of \STF\ with various stoichiometry ($x=0$, 0.04, 0.1, 0.12, 0.16, 0.19, 0.21, 0.25, 0.28, 0.44, 0.78 and 1) were prepared by standard electrochemical oxidation \cite{Geiser91}. Both BEDT-TTF and BEDT-STF were synthesized at Hokkaido University in Sapporo, where also the crystal growth is carried out. For the alloying series, the amount of donor molecules was preselected; for each batch the actual substitution value $x$ was determined {\it a posteriori} by energy-dispersive x-ray spectroscopy: using \Cu\ as a reference we compared the intensity of S atoms to that of Se atoms \cite{Saito18}.

\begin{table}[b]
\caption{Room-temperature values of the unit cell parameters of \Cu\ and \stf\ assuming space groups P2$_1$/c and P$\bar{1}$ for the analysis of the x-ray scattering results. Data for the former compound are taken from Refs.~\onlinecite{Pinteric14} and \onlinecite{Foury18}. \label{tab:unitcell}}
\begin{center}
\begin{tabular}{lcc|cc}
 & \multicolumn{2}{c}{$\kappa$-(BEDT-TTF)$_2$Cu$_2$(CN)$_3$} & \multicolumn{2}{|c}{$\kappa$-(BEDT-STF)$_2$Cu$_2$(CN)$_3$}\\
\noalign{\smallskip}\hline\noalign{\smallskip}
& P2$_1$/c & P$\bar{1}$ & P2$_1$/c & P$\bar{1}$  \\
\noalign{\smallskip}\hline\noalign{\smallskip}
$a$ & 16.0920(4)\,\AA & 16.1221(10)\,\AA  &16.2965(8)\,\AA &16.5625(8)\,\AA  \\
$b$ &8.5813(2)\,\AA  &8.591(6)\,\AA &8.6082(5)\,\AA &8.6292(3)\,\AA \\
$c$ &13.3904(4)\,\AA &13.412(8)\,\AA &13.3985(6)\,\AA & 13.4025(5)\,\AA{} \\
$\alpha$  &90$^\circ$   &89.99(2)$^\circ$ &90$^\circ$& 90.00$^\circ$\\
$\beta$  &113.381(3)$^\circ$   & 113.43(2)$^\circ$ &113.1300(16)$^\circ$ &115.0115(14)$^\circ$\\
$\gamma$  &90$^\circ$& 90.01(2)$^\circ$ &90$^\circ$& 90.00$^\circ$\\
$V$ &1697.25\,\AA$^{3}$ &1704.46\,\AA$^{3}$& 1728.5\,\AA$^3$ &1735.87\,\AA$^3$\\
$Z$ &2    &2 & 2 &2
\end{tabular}
\end{center}
\end{table}
The structure consists of $bc$ layers of strongly dimerized BEDT-TTF or BEDT-STF molecules, with
each dimer oriented approximately perpendicular to its nearest neighbors (Fig.~\ref{fig:structure}). Overall it is assumed that
the compounds crystallize in monoclinic and centrosymmetric P2$_1$/c space group.
Based on an x-ray study of the parent compound Foury-Leylekian {\it et al.} recently suggested
a triclinic symmetry P$\bar{1}$ with two non-equivalent dimers in the unit cell  \cite{Foury18},
however, the charge imbalance among the sites is extremely weak. Upon sulfur substitution the unit cell volume slightly increases (less than $2\,\%$) in a linear fashion without any change in symmetry. In Tab.~\ref{tab:unitcell} we list the unit cell parameters for the limiting cases, assuming both, P2$_1$/c and P$\bar{1}$ space group symmetry.

Electrical transport was measured parallel to the $c$-axis from room temperature down to $T=1.8$~K by standard four-probe technique. For this, thin gold wires were attached by carbon paste. Furthermore, we measured the complex electrical impedance as a function of temperature and frequency in order to obtain the dielectric permittivity $\hat{\varepsilon} = \varepsilon_{1} + {\rm i} \varepsilon_{2}$.
Here, gold wires were attached to opposite crystal surfaces and the data recorded by an impedance analyzer in the frequency range from 40~Hz to 10~MHz covering temperatures down to $T=5$~K.
The applied ac voltage was set to 0.5~V, making sure that we operate in the Ohmic regime.

The optical properties of several single crystals have been studied by reflectivity measurements using standard Fourier-transform spectroscopy from the far-infrared up to the near-infrared range \cite{Dressel04}. The light was polarized along the two principal optical axes, i.e.\ $E \parallel b$ and $E \parallel c$, on as-grown surfaces. Besides regular in-plane experiments, we also probed
the polarization perpendicular to the conducting planes, a direction which couples to the most charge-sensitive
infrared-active intramolecular vibrational mode $\nu_{27}(b_{1u})$.
The sample was cooled down to $T = 10$~K by a helium-cooled optical cryostat.
In addition, the high-frequency properties (up to 35\,000~\cm) were determined by spectroscopic ellipsometry
at ambient condition. The optical conductivity was calculated via
Kramer-Kronig analysis using a constant reflectivity extrapolation at low
frequencies and temperatures for the Mott insulators, while a Hagen-Rubens behavior was assumed for
elevated temperatures and substitions $x>0.12$ where metallic properties prevail \cite{Elsasser12,Sedlmeier12}.

\section{DC Transport}
\begin{figure}
	\centering
	\includegraphics[width=0.9\columnwidth]{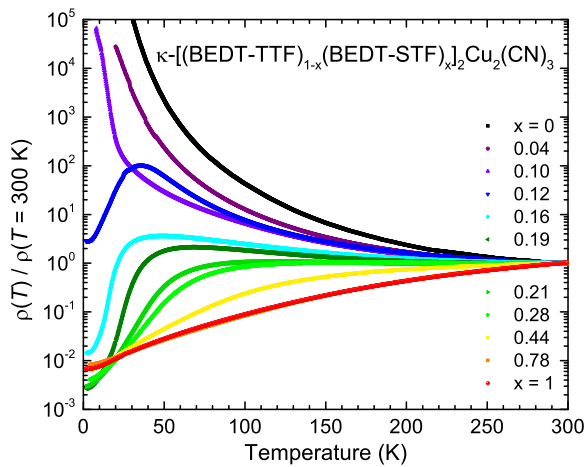}
	\caption{Temperature dependence of the dc resistivity of \STF\ for
various substitution values $x$ as indicated spanning the full range from the insulating $x=0$ to the metallic side. The data are measured along the highly conducting $c$-axis and normalized to the respective room temperature value for better comparison. }
	\label{fig:dc}
\end{figure}

Fig.~\ref{fig:dc} displays the $c$-axis dc resistivity $\rho(T)$ as a function of temperature for all samples in our substitution series from $x=0$ to 1. The room-temperature values increase  from approximately $0.03~\Omega$cm for $x=1$ to around $0.5~\Omega$cm for $x=0$. For $x=0.12$ and higher the systems turns metallic at low temperatures; the temperature range of metallic conductivity
increases for larger substitution and exceeds $T= 300$~K for $x\geq0.44$. The maximum in $\rho(T)$ defines the Brinkman-Rice temperature $T_{\rm BR}$ that increases upon alloying.
For very low temperatures, the metallic properties are clearly characterized by a $\rho(T)\propto T^2$ behavior that is the hallmark of electron-electron interaction. A detailed analysis of the Fermi-liquid properties will be provided elsewhere \cite{Pustogow20}.

For $x=0.1$ and 0.12 we identify traces of superconductivity in $\rho(T)$  that is suppressed by a strong magnetic field ($B=5$~T). Measurements of the magnetic susceptibility $\chi(T)$ confirm a superconducting transition temperature $T_c \approx 2.8$~K. A more detailed characterization of the superconducting state will be subject of a separate publication.

\section{Optical Properties}
\begin{figure}
	\centering
	\includegraphics[width=1\columnwidth]{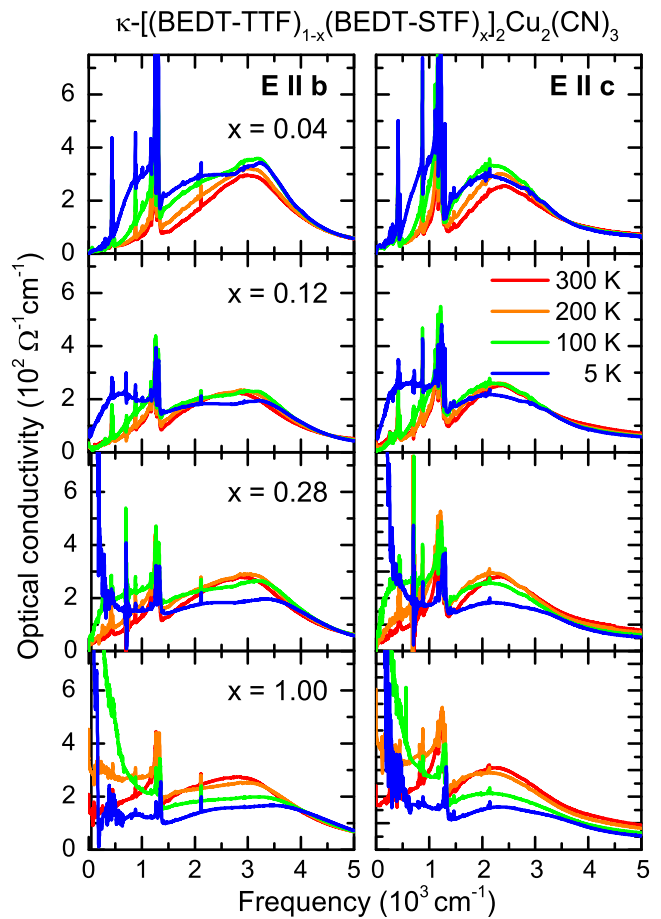}
	\caption{Frequency dependent conductivity of \STF\ for
several substitution values $x = 0.04$, 0.12, 0.28, and $x=1$ recorded for the two polarizations within the plane
(left column $E\parallel b$; right column $E\parallel c$)
at different temperatures as indicated.}
	\label{fig:Cond}
\end{figure}

The in-plane optical conductivity $\sigma_1(\omega)$ of \STF\ is displayed in Fig.~\ref{fig:Cond} for crystals of varying substitution $0\leq x \leq 1$. The data are measured for the in-plane directions $E\parallel b$ and $E \parallel c$ from room temperatures down to $T=5$~K. The properties are rather similar for both polarizations indicating electronically isotropic behavior within the $bc$-plane.

For BEDT-STF substitutions $x<0.1$, the compounds remain insulating at all temperatures;
but no Mott gap develops upon cooling. On the contrary,
similar to the pristine crystals \cite{Kezsmarki06,Elsasser12,Pustogow18b},
a pronounced in-gap absorption is present that becomes enhanced as or $T$ is reduced or $x$ increases.
In these cases we enter the coexistence regime of metallic and insulating regions expected for first-order phase transitions: the rising metallic fraction increasingly contributes to the optical properties. Eventually the alloys exhibit a Drude-like contribution to the optical conductivity, indicating the metallic properties in accord with the dc results presented in Fig.~\ref{fig:dc}.
An extended Drude analysis of the optical conductivity be presented elsewhere \cite{Pustogow20}, where the frequency dependence of the scattering rate $1/\tau$ and effective mass is extracted and the dependence on the substitutional value $x$ is discussed. According to the $\rho(T) \propto T^2$ behavior in the resistivity, we can identify a $1/\tau(\omega) \propto \omega^2$ dependence at the lowest temperatures.

\begin{figure}
	\centering
	\includegraphics[width=0.9\columnwidth]{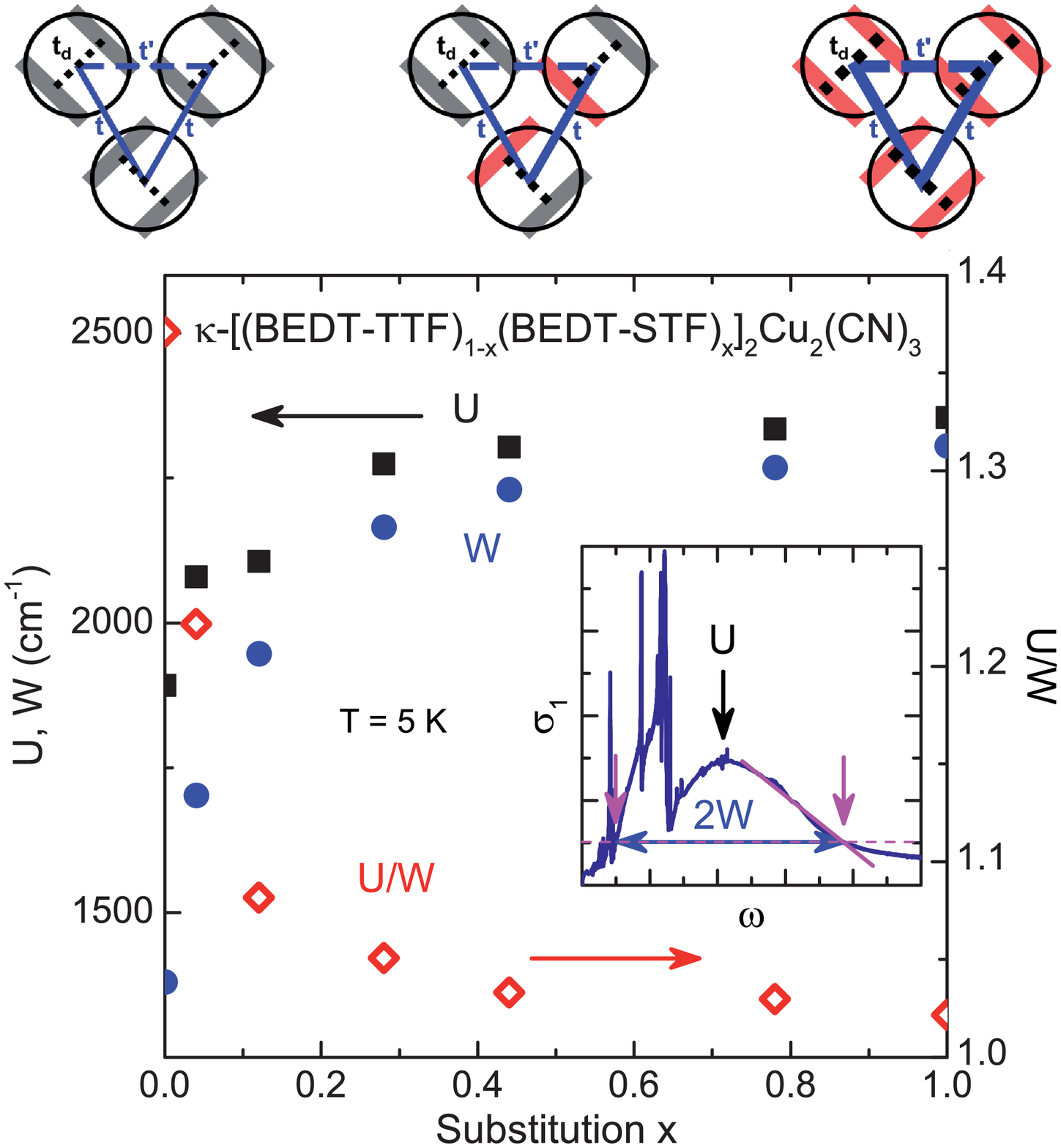}
	\caption{Dependence of the Coulomb interaction $U$ (full black squares) and bandwidth $W$ (solid blue dots) on the substitution $x$ in \STF\ as determined from low-temperature optical conductivity. The inset explains the determination of
$U$ from the mid-infrared maximum in $\sigma_1(\omega)$ and $W$ from the half-width at half maximum.
The effective correlation strength is given by the ratio $U/W$ (open red diamonds) and also plotted, corresponding to the right-hand scale. The three sketches above illustrate how the substitution $x$ gradually tunes \Cu\ to \stf; the lines of increasing thickness indicate the enhancement of transfer integrals $t$ (solid blue), $t^{\prime}$ (dashed blue) and $t_d$ (dotted black).}
	\label{fig:U-W}
\end{figure}

In Fig.~\ref{fig:Cond} we also see that the mid-infrared peak consists of two contributions: inter-dimer excitations and transitions between the  lower and the upper Hubbard bands, which almost coincide when probed along the $c$-axis. This has been observed in most $\kappa$-phase BEDT-TTF compounds and confirmed by {\it ab-initio} calculations \cite{Faltermeier07,*Merino08,*Dumm09,*Dressel09,Elsasser12,Ferber14}.
Following the approach taken in Ref.~\onlinecite{Pustogow18b}, we focus on the low-temperature spectra ($T=5$~K, $E\parallel c$) where we assign the maximum of the mid-infrared peak to the transitions between the lower and the upper Hubbard band separated by $U$. The width of the conductivity band in $\sigma_1(\omega)$ corresponds to twice the electronic bandwidth $W$, as illustrated in the inset of Fig.~\ref{fig:U-W} \footnote{Due to the metallic contribution at low frequencies, we determined the bandwidth as $1/2$ of the full width of the Mott-Hubbard band after subtracting a Drude fit for the compounds with metallic ground state. In particular, the band edge at the high-frequency tail $\omega_{\rm high}$ was defined from a linear cut-off; continuing $\sigma_1(\omega_{\rm high})$ to the low-frequency tail yields $\omega_{\rm low}$. Using these assignments, $2W$ corresponds simply to $\omega_{\rm high}-\omega_{\rm low}$. As a matter of course, the resulting Mott-Hubbard band parameters of CuCN are slightly different than in Ref.~\onlinecite{Pustogow18b}. Also $U$ was extracted after subtracting a low-frequency Drude fit, the frequency-dependent background of which slightly shifts the maximum of $\sigma_1(\omega)$ to higher energy compared to the raw data.}.

The extracted absolute values of $U$ and $W$ are plotted in the main frame as a function of STF-substitution $x$. We observe an increase of the electronic repulsion $U$ corresponding to the larger intra-dimer overlap $t_d$ as orbitals extend; even more pronounced is the rise of the bandwidth $W$, resulting from the increase of the transfer integrals $t$ and $t^{\prime}$ between adjacent dimers. The effect of correlations is measured by the ratio $U/W$, which is also plotted in Fig.~\ref{fig:U-W}, corresponding to the right scale. Obviously, the metallic compound \stf\ is much less correlated compared to the Mott insulator \Cu. The variation, however, is far from linear: a dramatic drop of $U/W$ is found for $x<0.2$, while for larger $x$
the ratio saturates around $U/W\approx 1.02$~\footnote{The apparent leveling off of $U/W$ around a value near unity
can be also a result of the method applied to determine the bandwidth. If the Hubbard bands strongly overlap, the simple scheme depicted in the inset of Fig.~\ref{fig:U-W} may underestimate $W$.}. Overall, from these optical experiments the Mott insulator-to-metal transition can be located around $x\approx 0.1$ in agreement with the dc resistivity.

\section{Vibrational Spectroscopy}
By now, it is well established that most of the dimerized BEDT-TTF salts, such as \Cl, \Cu, or \Ag\ do not possess sizeable charge disproportionation within the dimers \cite{Sedlmeier12,Pinteric16,Tomic15}, with a few exceptions, like $\kappa$-(BEDT-TTF)$_2$\-Hg(SCN)$_2$Cl \cite{Drichko14,*Lohle16,*Ivek17}. Raman and infrared spectroscopies are the best suited tools to locally probe the charge per molecule by inspecting the intramolecular vibrations that are known to be extremely sensitive to the amount of charge residing on the molecule \cite{Dressel04,Yamamoto05,Girlando11,*Girlando12}. Here we addressed the question whether charge imbalance occurs when sulfur is replaced by selenium in the alloys \STF.

\begin{figure}
	\centering
	\includegraphics[width=0.7\columnwidth]{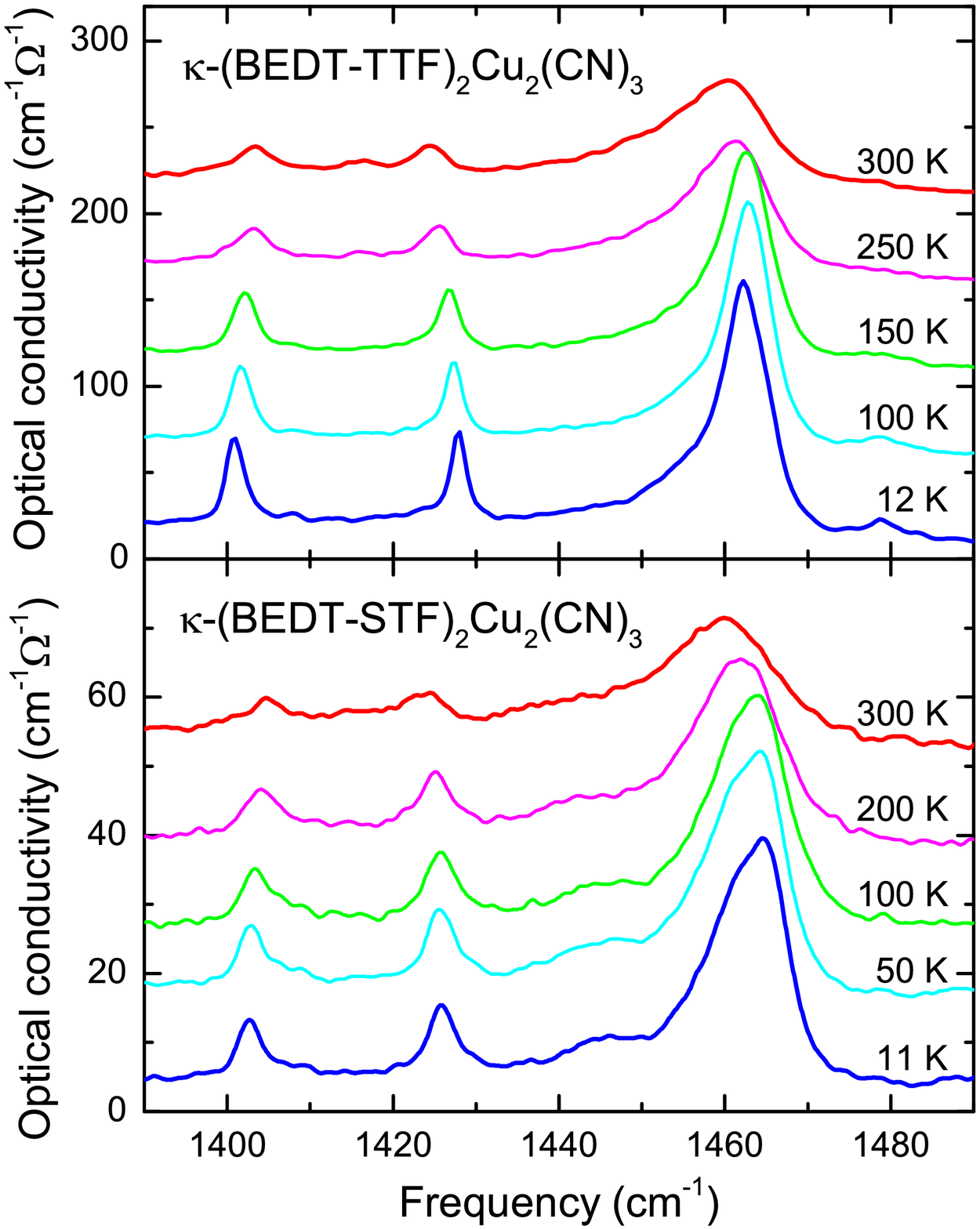}
	\caption{Optical conductivity of \Cu\ and \stf\ recorded for the out-of-plane polarization $E\parallel a$. The curves for the different temperatures are displaced for clarity reasons. The strong C=C vibrational mode $\nu_{27}(B_{1u})$ around 1460~\cm\ sharpens upon cooling but does not exhibit any severe modifications or splitting, as expected if strong charge disproportionation would occur. The data for \Cu\ were previously discussed in Ref.~\onlinecite{Sedlmeier12}.}
	\label{fig:nu27}
\end{figure}

To that end we recorded the optical properties perpendicular to the planes. In Fig.~\ref{fig:nu27} the out-of-plane conductivity of \Cu\ is compared to \stf\ in the finger-print spectral region where the most charge-sensitive molecular vibrations of the central C=C occur.
The dominant mode $\nu_{27}(B_{1u})$ alters only slightly upon STF-substitution. If we focus on the limiting cases $x=0$ and $x=1$, we see that both compounds exhibit basically the same temperature dependence: a typical sharpening and slight hardening upon cooling. This behavior is explained by the fact that the sufur/selenium atoms in the pentagonal inner ring are basically not involved in these modes.
It can best be seen from the vibrational analysis and animations of Ref.~\onlinecite{Dressel16} that these vibrations are well confined to the C=C bonds. Most important, since we do not observe any pronounced mode splitting in the \STF\ series,
we conclude the absence of any significant charge imbalance among the donor molecules.

\section{Dielectric Properties}
\begin{figure}
	\centering
	\includegraphics[width=1\columnwidth]{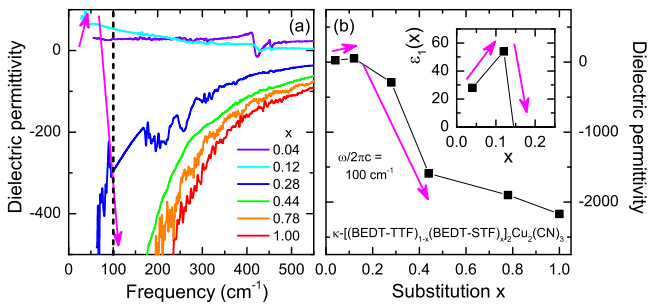}
	\caption{Despite continuously increasing low-frequency conductivity, in the low-frequency limit the dielectric permittivity $\varepsilon_1(x)$ exhibits a peak around the Mott transition, reminiscent of a percolative-type coexistence of metallic and insulating regions. The left panel displays the real part of the dielectric permittivity $\varepsilon_1(x)$ as obtained from far-infrared reflectivity measurements  for $E\parallel c$ at $T=5$~K for \STF\ with various substitutions $x$ as indicated. To better follow the substitutional dependence, panel (b) displays the dielectric permittivity taken at $\omega/2 \pi c = 100$~\cm\ indicated by the dashed line in panel (a).}
	\label{fig:permittivity}
\end{figure}

From the Kramers-Kronig analysis of the measured reflectivity spectra we obtain the complex dielectric constant $\hat{\varepsilon}(\omega, T, x)$ as a function of frequency, temperature and substitution.
Fig.~\ref{fig:permittivity} shows the real part $\varepsilon_1(\omega)$ for the different \STF\ crystals recorded at $T=5$~K. In the Mott insulating state ($x\leq 0.1$), the permittivity is basically frequency-independent and acquires a small, positive value. As $x$ increases, the quasi-static $\varepsilon_1(\omega \rightarrow 0)$ first increases before it rapidly drops to large negative values.
After crossing the Mott insulator-to-metal transition the system becomes conductive:
the strong screening of the coherent quasiparticle drives $\varepsilon_1$ negative.
A similar observation was reported for the Mott transition of VO$_2$, where the low-frequency permittivity diverges as a function of temperature. Near-field optical microscopy revealed that this behavior stems from the phase coexistence of metallic puddles in an insulating matrix~\cite{Qazilbash07}.
In general, the divergency of the dielectric permittivity $\varepsilon_1(x)$ is a hallmark of percolative phase transitions in microemulsions \cite{Grannan81,vanDijk86,Clarkson88a,*Clarkson88b,Alexandov99},
composites \cite{Pecharroman00,*Pecharroman01,Nan10} or percolating metal films \cite{Berthier97,Hovel10,*Hovel11,deZuani14,*deZuani16}.

\begin{figure}
	\centering
	\includegraphics[width=1\columnwidth]{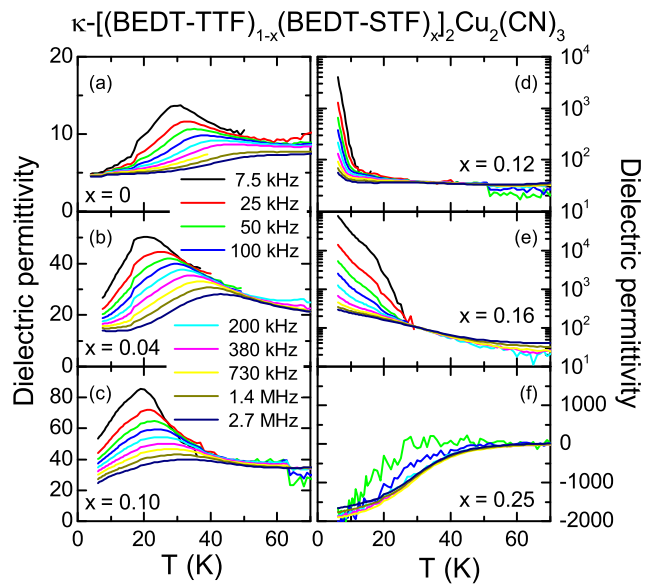}
	\caption{Temperature-dependent dielectric permittivity of \STF\ for substitutional values $x=0$, 0.04, 0.10, 0.12, 0.16 and 0.25 measured at several frequencies. Note the different ordinates used in the various panels. (a)~The pure crystal
exhibits a relaxor-type ferroelectric feature below $T=50$~K,  which becomes more pronounced and shifts to lower $T$ as frequency gets smaller. (b,c)~As $x$ is increased to 0.1, $\varepsilon_{1}(T)$ rises strongly and the peak appears at lower temperatures.
(d-e)~Eventually the permittivity reaches values of $10^5$ due to the coexistence of spatially separated metallic and insulating
regions. The response is strongly frequency dependent.
(f)~Upon percolation around $x=0.2$, the dielectric constant is negative,
giving evidence for the metallic behavior that continues for all higher substitutions up to $x=1$.}
	\label{fig:eps1}
\end{figure}

Since audio- and radio-frequency experiments are more suitable for exploring the dielectric behavior at the insulator-metal transition, we have conducted dielectric experiments down to 40~Hz. Fig.~\ref{fig:eps1} summarizes the dielectric response of \Cu\ and how it is affected by moving across the Mott insulator-to-metal transition via STF-substitution. We plot the real part of the permittivity $\varepsilon_1$ as a function of temperature $T$ for selected frequencies $f$ and substitutions $x$, as indicated.
The pronounced peak dominating the temperature dependence of $\varepsilon_1(T)$ was discovered by Abdel-Jawad {\it et al.} \cite{Abdel10} and subsequently confirmed by other groups \cite{Pinteric14,Pinteric15,Tomic15,Sasaki15,Rosslhuber19}.
When probed at $f=7.5$~kHz, the maximum is observed around $T=30$~K in the case of \Cu\ [Fig.~\ref{fig:eps1}(a)]; with a slight sample-to-sample dependence, in agreement with previous reports. The peak shifts to higher temperatures as the frequency increases;
at the same time, however, it gets less pronounced. This behavior resembles the well-known phenomenology of relaxor ferroelectrics \cite{Cross08}.

Already the minimal substitution of $x=0.04$ and 0.1 enhances the dielectric permittivity significantly
with the maximum $\varepsilon_1 (f=7.5~{\rm kHz}) \approx 50$ and 80.
This strong increase of $\varepsilon_1 (T)$ and concomitant shift of the peak to lower
temperatures when we approach the insulator-metal transition
is in full accord with our pressure-dependent dielectric studies \cite{Rosslhuber19}
where an extensive and detailed analysis is given.
As we approach the phase transition further ($x=0.12$ and 0.16),
the dielectric constant drastically diverges,
reaching values up to $10^5$ in the static low-temperature limit.
A divergency in $\varepsilon_1(x)$ is the fingerprint of a percolative phase transition where metallic regions form in an insulating matrix \cite{Mott90,Kirkpatrick73,Efros76}. When crossing the percolation threshold, the system acts like a metal,
characterized by a negative dielectric permittivity, $\varepsilon_1 < 0$.
With rising $x$ the sign change of the dielectric constant traces the Brinckman-Rice temperature, as it was identified by the maximum in $\rho(T)$ (Fig.~\ref{fig:dc}) \cite{Pustogow20}. These results confirm the observations we extracted from the optical response in Fig.~\ref{fig:permittivity}.

\begin{figure}
	\centering
	\includegraphics[width=0.8\columnwidth]{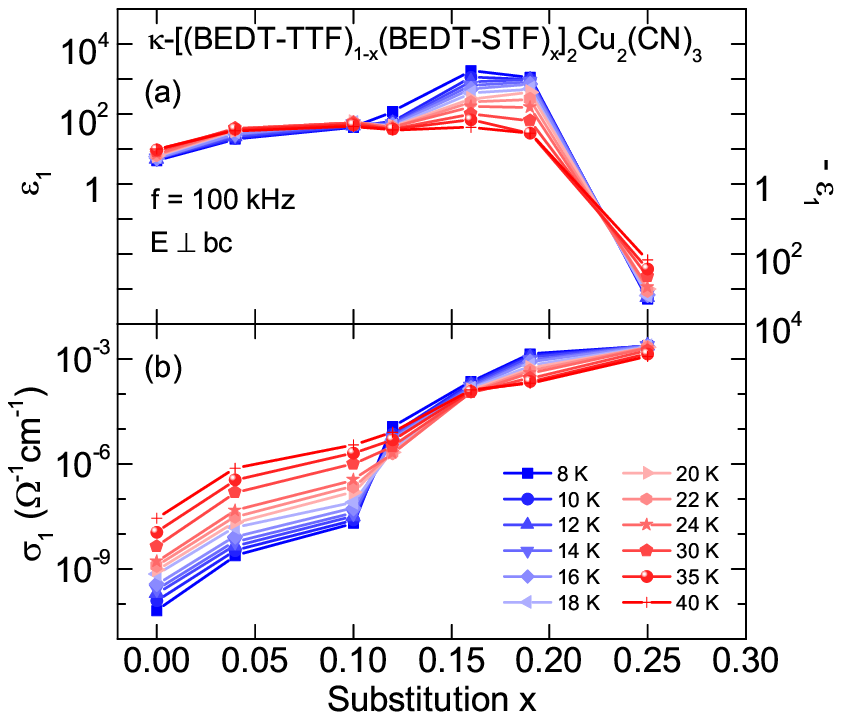}
	\caption{Dielectric properties of \STF\ in dependence of the substitution level $x$, recorded perpendicular to the $bc$-plane at  a fixed frequency $f = 100$~kHz at different temperatures as indicated. Note the logarithmic scales.
(a)~The permittivity $\varepsilon_1(x)$ forms a pronounced maximum followed by a rapid drop to negative values, corresponding to right scale. (b)~The conductivity $\sigma_1(x)$ exhibits a step-like increase at $x = 0.12$, indicating the Mott transition. The observed behavior matches the signature of a percolating system. At the insulator-to-metal transition, the nucleation and growth of metallic puddles sets in, which are spatially separated in an insulating matrix.
Upon increasing $x$, the metallic filling fraction grows until the metallic state is
completely established at $x = 0.25$. With rising temperature the features
diminish in amplitude or step size, respectively, consistent with the change
from the first-order insulator-metal transition to the crossover region at elevated temperatures. }
	\label{fig:eps_sig_x}
\end{figure}
For better comparison, we plot $\varepsilon_1(x)$ and $\sigma_1(x)$ for the complete series \STF\
in Fig.~\ref{fig:eps_sig_x} measured at various temperatures $T$.
Raising the STF-content results in a strong enhancement of the dielectric constant up to $10^3$, whereas for $x = 0.25$ a drop to large negative values of the order of
$\varepsilon_1(x) \approx -10^3$ is observed. Concurrently, $\sigma_1(x)$ increases by several orders of magnitude: the rapid rise at $x \approx 0.12$ indicates the crossing of the phase boundary into the metallic regime.

\section{Discussion and Summary}
In a complementary study \cite{Rosslhuber19} the dielectric properties of \Cu\ have been measured as a function of hydrostatic pressure. This is an alternative way to shift the Mott insulator through the first-order phase transition until the Fermi-liquid state is reached above $p \approx 1$~kbar.
The behavior obtained by the pressure-dependent investigations is in full accord with the dielectric response observed here in the substitutional series; both can be well described by dynamical mean-field theory on the single-band Hubbard
model when combined with percolation theory \cite{Pustogow19}.

\begin{figure}
	\centering
	\includegraphics[width=0.9\columnwidth]{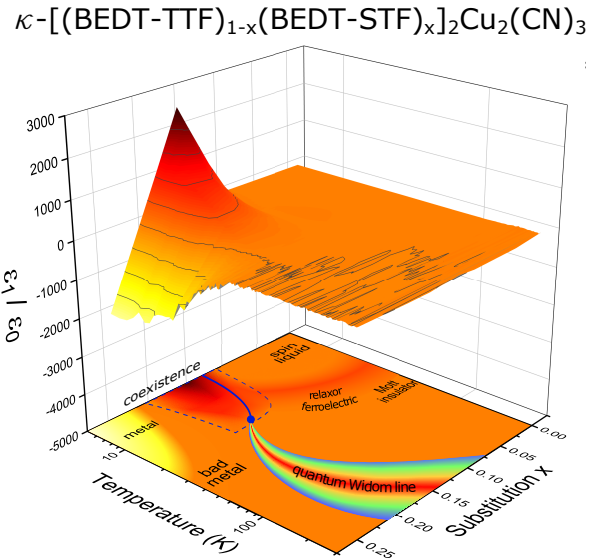}
	\caption{Contour plot of the dielectric permittivity $\varepsilon_{1}$ recorded at $f=380$~kHz as a function of temperature and STF-substitution $x$. We observe a strong increase of $\varepsilon_{1}$ up to 2500, centered around $x=0.15$-0.2 and below $T\approx 15$-20~K, close to the first-order Mott transition. We ascribe this to phase coexistence around the first-order Mott transition hosting spatially separated metallic and insulating regions, in agreement with state-of-the-art dynamical-mean-field-theory \cite{Vucicevic13,Pustogow19}. See text for more details.}
	\label{fig:Contour}
\end{figure}
The fact that both approaches reach very much the same results provides firm evidence for a percolative Mott transition.
Substituting BEDT-STF molecules increases the intra- and interdimer transfer integrals on a local scale, but this effect is smeared out.
Local strain does not alter the crystal symmetry; the substitution results in a slightly larger unit cell as seen from Tab.~\ref{tab:unitcell}. Since the organic molecules are rather large, diffusion within the crystal is prohibited, ruling out the formation of BEDT-STF clusters or domains of extended size beyond the stochastic occurrence. As depicted in Fig.~\ref{fig:U-W}, STF-substitution only results in a gradual decrease of effective correlation strength $U/W$.

Our findings are summarized in Fig.~\ref{fig:Contour} where the dielectric permittivity $\varepsilon_1$ is plotted as a function of temperature and STF-substitution $x$ producing a contour plot that constitutes the corresponding phase diagram.
We conclude that the discovery of the enormous divergency in the low-temperature quasi-static dielectric permittivity $\varepsilon_1(x)$ proves metallic regions coexisting with the insulating matrix.
The reduction  of electronic interactions $U/W$ due to increasing $x$ leads to a spatial phase separation that is confined to temperatures below the critical endpoint $T_{\rm crit} \approx 15$-20~K. With rising metallic fraction, the resistivity decreases, as can be well understood by effective medium models  \cite{Choy15}. At large substitutions, metallic fluctuations dominate the electronic response. This range is well distinct by the negative and very large permittivity; its boundary trails the Brinkman-Rice temperature for the onset of metallic conductivity.
The complete electrodynamical properties of this correlated electron system including the
percolative aspect of the first-order Mott transition can be reproduced by the hybrid approach
of dynamical mean-field theory amended with percolation theory \cite{Pustogow19}.
Finally we would like to point out that the novel path of tuning the effective correlations by increasing the molecular orbitals
when partially replacing BEDT-TTF by the selenium-containing BEDT-STF molecules can be applied to other charge-transfer salts as well, providing an interesting alternative to chemical and physical pressure.

\section*{Acknowledgements}
We acknowledge continuous discussions with V. Dobrosavljevi\'c and Y. Tan.
We thank N. Kircher for participating in some experiments and Y. Takahashi for performing x-ray diffraction;
G. Untereiner put an enormous efforts into sample preparation.
The project was supported by the Deutsche Forschungsgemeinschaft (DFG) via DR228/39-3.

%

\end{document}